\documentstyle[preprint,aps,epsfig]{revtex}

\begin{document}
\draft
\title {Optical nonlinearity enhancement of graded \\ metal-dielectric composite films}
%\date {\today}
\author {J. P. Huang$^{1,2}$, L. Dong$^{1,3}$, K. W. Yu$^1$}
\address {$^1$Department of Physics, The Chinese University of Hong Kong, 
 Shatin, NT, Hong Kong \\
 $^2$Max Planck Institute for Polymer Research, Ackermannweg 10, 
 55128 Mainz, Germany\\
$^3$Biophysics and Statistical Mechanics Group,
Laboratory of Computational Engineering,
Helsinki University of Technology,
P.\,O.~Box 9203, FIN-02015 HUT, Finland }

\maketitle

\begin{abstract}

We have derived the local electric field inside graded 
metal-dielectric composite films with weak nonlinearity analytically,  which further yields the effective linear dielectric constant and   
third-order nonlinear susceptibility of the graded structures.
As a result, the composition-dependent gradation can produce a broad resonant plasmon band in the optical region,
resulting in a large enhancement of the optical nonlinearity and hence
a large figure of merit.

\vskip 5mm
\pacs{PACS number(s): 77.22.Ej, 42.65.An, 42.79.Ry}
\end{abstract}

%\section{Introduction}

%\section{Formalism}

In contrast to bulk materials, the corresponding thin films often possess different optical properties~\cite{Grull,Kammler}. It is also known that graded materials~\cite{Milton,Jones} have quite different physical properties from the 
homogeneous materials~\cite{Yamanouchi}.  To one's interest, one found that the graded thin films may have better dielectric 
properties than a single-layer film~\cite{LuAPL03}. 
However, the traditional theories~\cite{Jackson,AIP} fail to deal with the
composites of graded particles. For this purpose, we put forth a first-principles approach~\cite{Dong,Gu-JAP} 
and a differential effective dipole approximation~\cite{Huang}. The situation becomes more complicated by the presence of nonlinearity which exists in 
real composites~\cite{Stroud,Agarwal,Zeng,DJBergmanPRB89,Scaife,Bergman,YuPRB93,Shalaev,Hui,Gao1,Tlidi}.
Moreover, there is a great need for nonlinear optical materials with large 
nonlinear susceptibility or optimal figure of merit (FOM). Therefore, much work has been done on how to gain a large 
nonlinearity enhancement or optimal FOM of bulk composites by the 
surface-plasmon resonance in metal-dielectric 
composites~\cite{Shalaev}, as well as by taking 
into account the structural information~\cite{Sipe,Gao3}. 
Recently, a large nonlinearity enhancement was found for a sub-wavelength multilayer of titanium dioxide and 
conjugated polymer~\cite{Fischer}.

However, the surface plasmon resonant nonlinearity enhancement often occurs concomitantly with a strong absorption, and unfortunately this behavior renders the FOM 
of the resonant enhancement peak to be too small to be useful. 
To circumvent this 
problem, we shall consider a kind of graded metal-dielectric composite film, in which  a dielectric component is introduced as 
spherical particles embedded in the metallic component.

%We have found that the composition-dependent gradation  in the
%films yields a broad resonant plasmon band in the optical region.
%Thus, we have succeeded in gaining a large nonlinearity enhancement as well as 
%optimal FOM by considering the effect of positional dependence of gradation.

Let us start by considering a graded  film with width $L$. Here the gradation of interest is in the direction perpendicular to the film. In this connection,  the local 
constitutive relation between the displacement ${\bf D}$ and  
electric field ${\bf E}$  is given by
\begin{equation}
{\bf D}(z,\omega) = \epsilon(z,\omega){\bf E}(z,\omega)+\chi(z,\omega)
 |{\bf E}(z,\omega)|^2{\bf E}(z,\omega),
\label{Weak} 
\end{equation}%1
where $\epsilon (z,\omega)$ and $\chi(z,\omega)$ are respectively the 
linear dielectric constant and third-order nonlinear susceptibility of a certain layer inside the graded film. 
It is worth noting that both $\epsilon(z,\omega)$ and $\chi(z,\omega)$ are gradation 
profiles as a function of position $z$. In view of the metal-dielectric composite layer,   $\epsilon (z,\omega)$ 
 is given by the well-known Maxwell-Garnett approximation
\begin{equation}
\frac{\epsilon (z,\omega)-\epsilon_1(\omega)}{\epsilon (z,\omega)+2\epsilon_1(\omega)}=p(z)\frac{\epsilon_2-\epsilon_1(\omega)}{\epsilon_2+2\epsilon_1(\omega)},\label{MG}
\end{equation}
where $\epsilon_2$ stands for the dielectric constant of the dielectric, and $p(z)$ the layer dielectric profile. It is worth noting that Eq.~(\ref{MG}) gives an approximate
$\epsilon(z,\omega)$ for the present geometry.  Here the dielectric function of the metal $\epsilon_1(\omega)$ is given by the Drude expression
\begin{equation}
\epsilon_1(\omega) = 1-\frac{\omega_p^2}{\omega(\omega+i \gamma)},
\end{equation}
where $\omega_p$ denotes the bulk plasmon  frequency, and $\gamma$ the collision frequency. Regarding Eq.~(\ref{MG}), we should remark more. In fact, it is not possible to calculate  $\epsilon (z,\omega)$  exactly in terms of the layer dielectric profile $p(z)$. However, to obtain an estimate of  $\epsilon (z,\omega),$ we can take a small volume element inside the layer, at a position $z$. Further, this small volume element can be seen as a composite where the dielectric particles are randomly embedded in the metallic component. Accordingly, the volume fraction of the dielectric particles is $p(z)$. In this regard, the Maxwell-Garnett approximation [namely, Eq.~(\ref{MG})] holds well for computing  $\epsilon (z,\omega)$.

Let us further assume that the weak nonlinearity condition is satisfied in the present work. 
That is, the contribution of the second term (nonlinear part 
$\chi(z,\omega)|{\bf E}(z,\omega)|^2$) in the right-hand side of 
Eq.~(\ref{Weak}) is much less than that of the first term 
(linear part $\epsilon(z,\omega)$)~\cite{Stroud}. 
Next, we focus on the quasi-static approximation, 
under which the whole graded structure can be regarded as an effective 
homogeneous one with effective (overall) linear dielectric constant 
$\bar{\epsilon}(\omega)$ and effective (overall) third-order nonlinear 
susceptibility $\bar{\chi}(\omega) .$ In the mathematical expression, the definition of  
$\bar{\epsilon}(\omega)$ and $\bar{\chi}(\omega)$ is given by~\cite{Stroud} 
\begin{equation}
\langle{\bf D}\rangle = \bar{\epsilon}(\omega){\bf E}_0
 +\bar{\chi}(\omega)|{\bf E}_0|^2{\bf E}_0, 
\end{equation}%2
where $\langle\cdots\rangle$ stands for the spatial average of $\cdots$, and 
${\bf E}_0 = E_0\hat{e}_z$  the applied field along  $z-$axis.

Owing to the simple graded structure, we can use the 
equivalent capacitance of series combination to calculate the linear 
response (i.e., optical absorption for the metallic film),
\begin{equation}
{1\over \bar{\epsilon}(\omega)} = {1\over L}
 \int_0^L {{\rm d}z\over \epsilon(z,\omega)}.
\end{equation}%6
For  calculating the nonlinear optical response, we first calculate local electric field $E(z,\omega)$ by the identity
\begin{equation}
\epsilon(z,\omega) E(z,\omega)=\bar{\epsilon}(\omega) E_0
\end{equation}
due to the virtue of the continuity of the electric displacement.
In view of the existence of nonlinearity inside the graded film, 
the effective nonlinear response $\bar{\chi}(\omega)$ can be given by~\cite{Stroud} 
\begin{equation}
\bar{\chi}(\omega){\bf E}_0{}^4 = \langle\chi(z,\omega)
 |{\bf E}_{{\rm lin}}(z)|^2{\bf E}_{{\rm lin}}(z)^2 \rangle , 
\end{equation}%7
where $E_{{\rm lin}}$ denotes the linear local electric field.
We can take one step forward to express the effective nonlinear response as an integral 
over the film,
\begin{equation}
\bar{\chi}(\omega) = {1\over L} \int_0^L {\rm d}z \chi(z,\omega)
 \left|{\bar{\epsilon}(\omega) \over \epsilon(z,\omega)}\right|^2
\left({\bar{\epsilon}(\omega) \over \epsilon(z,\omega)}\right)^2 .\label{Chi-8}
\end{equation}%8
In fact, the assumption of a $z$-dependent Maxwell-Garnett type dielectric profile is
equivalent to assume that the local field depends on $z$ only, see Eq.~(\ref{MG}). However, the real $\bar{\chi}(\omega)$ should
involve an integral over $x$, $y$, and $z$ of the local $\chi(x, y, z, \omega)$ multiplied
by terms involving $\epsilon(x, y, z, \omega)$. Thus, Eq.~(\ref{Chi-8}) offers an
approximate $\bar{\chi}(\omega)$, as expected.
%\section{Numerical results}

In what follows, we shall do some numerical calculations. First, set $\chi(z,\omega)$ to be a constant 
$\chi_1$. By doing so, the two components are assumed to
have the same real and positive frequency-independent
$\chi_1$,  so that we could emphasize the enhancement of the optical nonlinearity.   Regarding the layer dielectric profile, let us take a power former $p(z)=az^m$.
Without loss of generality, the layer width $L$ is taken to be $1$. 

In Fig.~1, we display (a) the optical absorption
$\sim\mathrm{Im}[\bar{\epsilon}(\omega)] ,$ (b) the modulus of the
effective third-order optical nonlinearity enhancement
$|\bar{\chi}(\omega)|/\chi_1 ,$ and (c) the FOM 
$|\bar{\chi}(\omega)|/\{\chi_1\mathrm{Im}[\bar{\epsilon}(\omega)]\}$
as a function of the incident angular frequency $\omega .$ 
Here $\rm{Im}[\cdots]$ means the imaginary part of $\cdots .$  To one's interest, when the layer dielectric profile $p(z)$ is taken into account, a broad resonant plasmon band is observed always. In other words, the broad band is caused to appear by the effect of the positional dependence of the dielectric or metal. Also, we  find that increasing $a$    causes the resonant band not only to be enhanced, but also red-shifted (namely, located at a lower frequency region). In this work $a$ should satisfy $0<a<1$, in order to ensure a $p(z)$ which is always smaller than unity.      In a word, although the enhancement of the effective third-order optical nonlinearity is often accompanied with the appearance of the optical absorption, the FOM is still possible to be very attractive due to the presence of the positional dependence of the dielectric or metallic components. Moreover, it is worth noting that  a
prominent surface plasmon resonant peak appears at somewhat higher
frequencies in addition to the surface plasmon band. As $a$ increases, this peak is blue-shifted (i.e., locates at a higher frequency region) accordingly.

Similarly, Figure~2 displays the influence of $m.$ It is apparent to see that the broad resonant plasmon band can be enhanced significantly by adjusting $m$. However, no distinct red-shift occurs for the plasmon band as $m$ varies. In contrast, we notice that increasing $m$ can make the surface plasmon resonant peak red-shifted.

As a matter of fact, the present results do not depend crucially on the
particular form of the layer dielectric profile $p(z)$. The only requirement is that we must have a  composition-dependent  layer to yield a broad plasmon band for the graded film.
It should be remarked that the optical response of the graded structure 
depends on polarization of the incident light, because the incident 
optical field can always be resolved into two polarizations.
However, a large nonlinearity enhancement occurs only when the electric 
field is parallel to the direction of the gradient~\cite{Fischer}, and 
the other polarization does not produce nonlinearity enhancement at 
all~\cite{Fischer}. Fortunately, the nonlinear susceptibilities of both the parallel and
perpendicular polarizations are related to the nonlinear phase shift,
which can be measured by using the z-scan method~\cite{Fischer}.
It is of interest to extend our consideration to composites 
in which graded spherical particles are embedded in a host medium~\cite{HGYG} 
to account for mutual interactions among graded particles~\cite{Gao3}. 

To calculate  $\epsilon (z,\omega),$ we used Eq.~(\ref{MG}) in which metal (or dielectric) serves as a host (or inclusion). Inversely, we could see metal (or dielectric) as an inclusion (or host), and use the same form as Eq.~(\ref{MG})  by exchanging $\epsilon_1(\omega)$ and $\epsilon_2$, and $p(z)$ and $1-p(z)$. In case of $\epsilon_1(\omega)>\epsilon_2$, the former  always gives the upper bound, while the latter the lower bound, and vice versa. The exact result must lie between the two bounds~\cite{HS}.   

To sum up, we have studied a graded metal-dielectric composite film, and found that the composition-dependent gradation  in the
film yields a broad resonant plasmon band in the optical region.
Thus, it is possible to gain a large nonlinearity enhancement as well as 
optimal FOM for graded metal-dielectric composite films.

{\it Acknowledgments.}
This work was supported by the Research Grants Council of the Hong Kong 
SAR Government under project number CUHK 403303, and in part by the DFG under 
Grant No. HO 1108/8-3 (J.P.H.).

\begin{figure}[h]
\caption{(a) Linear optical absorption 
${\rm Im}[\bar{\epsilon}(\omega)]$, (b) enhancement of the
third-order optical nonlinearity $|\bar{\chi}(\omega)|/\chi_1$,
and (c)  FOM (figure of merit) $\equiv
|\bar{\chi}(\omega)|/\{{\chi_1\rm Im}[\bar{\epsilon}(\omega)]\}$
versus the normalized incident angular frequency
$\omega/\omega_p$ for layer dielectric profile $p(z)=az^m,$ for different $a$ at $m=1.0.$ Solid line: $a=0.2;$ Dashed line: $a=0.4;$ Long-dashed line: $a=0.6;$ Dot-dashed line $a=0.8.$
 Parameters: $\gamma/\omega_p=0.01$ and $\epsilon_2=(3/2)^2.$}
\end{figure}

\begin{figure}[h]
\caption{Same as Fig.1, but for different $m$ at $a=0.8$. Solid line: $m=0.2;$ Dotted line: $m=0.6;$ Dashed line: $m=1.0;$ Long-dashed line: $m=1.4;$ Dot-dashed line $m=1.8.$  }
\end{figure}

\newpage
\centerline{\epsfig{file=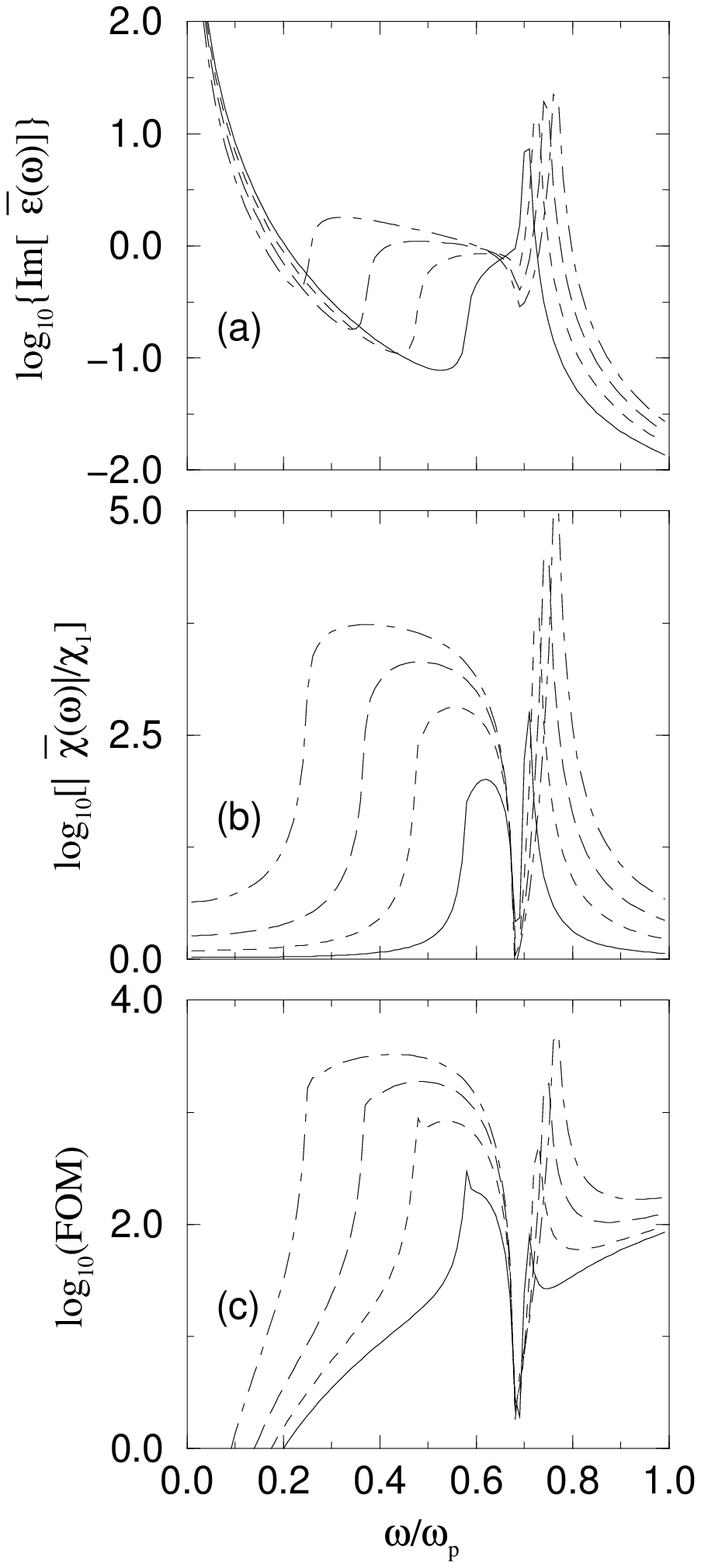,width=250pt}}
\centerline{Fig.~1./Huang,Dong,Yu}

\newpage
\centerline{\epsfig{file=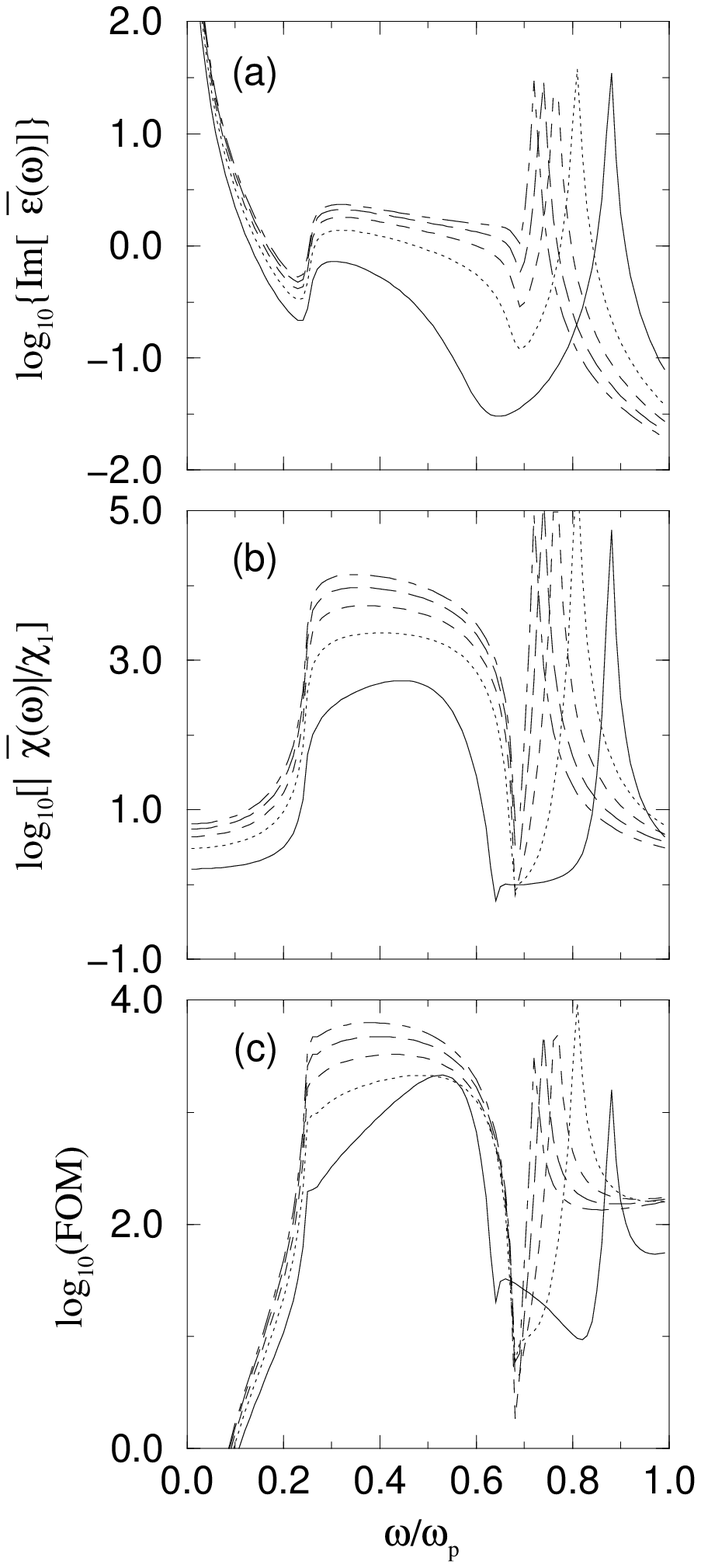,width=250pt}}
\centerline{Fig.~2./Huang,Dong,Yu}


\begin{references}

\bibitem{Grull} H. Grull, A. Schreyer, N. F. Berk, C. F. Majkrzak, and C. C. Han, Europhys. Lett. {\bf 50}, 107 (2000).


\bibitem{Kammler} D. R. Kammler, T. O. Mason, D. L. Young, 
 T. J. Coutts, D. Ko, K. R. Poeppelmeier, and D. L. Williamson, 
 J. Appl. Phys. {\bf 90}, 5979 (2001).



\bibitem{Milton} G. W. Milton, {\it The Theory of Composites},
 Chapter 7 (Cambrige University Press, Cambridge, 2002).

\bibitem{Jones} T. B. Jones, {\it Electromechanics of Particles} (Cambridge Univ. Press, Cambridge, 1995).

\bibitem{Yamanouchi} M. Yamanouchi, M. Koizumi, T. Hirai, and I. Shioda,
 in {\it Proceedings of the First International Symposium on
 Functionally Graded Materials} (Sendi, Japan, 1990).

\bibitem{LuAPL03} S. G. Lu, X. H. Zhu, C. L. Mak, K. H. Wong, 
 H. L. W. Chan, and C. L. Choy, Appl. Phys. Lett. {\bf 82}, 2877 (2003).




\bibitem{Jackson} J. D. Jackson, {\it Classical Electrodynamics}
 (Wiley, New York, 1975).
\bibitem{AIP} See, for example, the articles in
 {\it Electric Transport and Optical Properties of Inhomogeneous Media},
 edited by J. C. Garland and D. B. Tanner, AIP Conference Proceedings
 No.40 (AIP New York, 1978).
\bibitem{Dong} L. Dong, G. Q. Gu, and K. W. Yu, Phys. Rev. B {\bf 67},
 224205 (2003).
\bibitem{Gu-JAP} G. Q. Gu and K. W. Yu, J. Appl. Phys. {\bf 94}, 3376
 (2003).

%\bibitem{Yu1} K. W. Yu, G. Q. Gu and J. P. Huang, preprint:
% cond-mat/0211532.

\bibitem{Huang} J. P. Huang, K. W. Yu, G. Q. Gu, and M. Karttunen,
 Phys. Rev. E {\bf 67}, 051405 (2003).

\bibitem{Stroud} D. Stroud and P. M. Hui, Phys. Rev. B {\bf 37}, 
 8719 (1988).
\bibitem{Agarwal} G. S. Agarwal and S. Dutta Gupta, Phys. Rev. A
 {\bf 38}, 5678 (1988).
\bibitem{Zeng} X. C. Zeng, D. J. Bergman, P. M. Hui, and D. Stroud,
 Phys. Rev. B {\bf 38}, 10970 (1988).
\bibitem{DJBergmanPRB89} D. J. Bergman, Phys. Rev. B {\bf 39}, 4598 (1989).
\bibitem{Scaife} B. K. P. Scaife, {\it Principles of Dielectrics}
 (Calvendon Press, Oxford, 1989).

\bibitem{Bergman} D. J. Bergman and D. Stroud, {\it Solid State Physics:
 Applied in Research and Applications}, edited by H. Ehrenreich and
 D. Turnbull (Academic Press, New York, 1992), {\bf 46}, p. 147.

\bibitem{YuPRB93} K. W. Yu, P. M. Hui, and D. Stroud, Phys. Rev. B 
 {\bf 47}, 14150 (1993).
\bibitem{Shalaev} V. M. Shalaev, {\it Nonlinear Optics of Random Media:
 Fractal Composites and Metal-Dielectric Films}
 (Springer-Verlag, Berlin, 2000).
\bibitem{Hui}  See, for example, the articles in
 {\it Proceedings of the Fifth International Conference on Electrical
 Transport and Optical Properties of Inhomogeneous Media}, edited by 
 P. M. Hui, Ping Sheng, and L. H. Tang [Physica A {\bf 279}, 2000].

%\bibitem{Sarychev} A. K. Sarychev and V. M. Shalaev, Phys. Rep.
% {\bf 335}, 275 (2000).

\bibitem{Gao1} L. Gao, J. T. K. Wan, K. W. Yu, and Z. Y. Li, 
 J. Appl. Phys. {\bf 88}, 1893 (2000).

\bibitem{Tlidi} M. Tlidi, M. F. Hilali, and P. Mandel, Europhys. Lett. {\bf 55}, 26 (2001).


\bibitem{Sipe} J. E. Sipe and R. W. Boyd, Phys. Rev. A {\bf 46}, 
 1614 (1992).
\bibitem{Gao3} L. Gao, K. W. Yu, Z. Y. Li, and B. Hu, Phys. Rev. E
 {\bf 64}, 036615 (2001).
\bibitem{Fischer} G. L. Fischer, R. W. Boyd, R. J. Gehr, S. A. Jenekhe,
 J. A. Osaheni, J. E. Sipe, and L. A. Weller-Brophy,
 Phys. Rev. Lett. {\bf 74}, 1871 (1995).



\bibitem{HGYG} J. P. Huang, L. Gao, K. W. Yu, and G. Q. Gu, Phys. Rev. E {\bf 69}, 036605 (2004).

\bibitem{HS} Z. Hashin and S. Shtrikman, J. App. Phys. {\bf 33}, 3125 (1962).

\end{references}
\end{document}